\documentclass[%
reprint,
superscriptaddress,
amsmath,amssymb,
aps
]{revtex4-2}

\usepackage{graphicx}
\usepackage[dvipsnames]{xcolor}
\usepackage{dcolumn}
\usepackage{bm}
\usepackage[utf8]{inputenc}
\usepackage[T1]{fontenc}
\usepackage{amsmath}

\draft 

\begin{document}

\title[Dong et al.]{Enhanced mobility of ternary InGaAs quantum wells through digital alloying}

\author{Jason T. Dong}
\affiliation{Materials Department, University of California, Santa Barbara, CA 93106}

\author{Yilmaz Gul}
\affiliation{London Centre for Nanotechnology, University College London, 17-19 Gordon Street, London WC1H
0AH, United Kingdom}

\author{Aaron N. Engel}
\affiliation{Materials Department, University of California, Santa Barbara, CA 93106}

\author{Teun A. J. van Schijndel}
\affiliation{Department of Electrical and Computer Engineering, University of California, Santa Barbara, CA 93106}

\author{Connor P. Dempsey}
\affiliation{Department of Electrical and Computer Engineering, University of California, Santa Barbara, CA 93106}

\author{Michael Pepper}
\affiliation{London Centre for Nanotechnology, University College London, 17-19 Gordon Street, London WC1H
0AH, United Kingdom}
\affiliation{Department of Electronic and Electrical Engineering, University College London, Torrington Place,
London WC1E 7JE, United Kingdom}

\author{Christopher J. Palmstr\o m}
\email[]{cjpalm@ucsb.edu}
\affiliation{Materials Department, University of California, Santa Barbara, CA 93106}
\affiliation{Department of Electrical and Computer Engineering, University of California, Santa Barbara, CA 93106}

\date{\today}

\begin{abstract}
High In content InGaAs quantum wells (In $\geq$ 75\%) are potentially useful for topological quantum computing and spintronics applications. In high mobility InGaAs quantum wells, alloy disorder scattering is a limiting factor. In this report, we demonstrate that by growing the InGaAs quantum wells as a digital alloy, or a short period superlattice, we can reduce the alloy disorder scattering within the quantum well and increase the peak 2 K electron mobility to 545,000 cm\textsuperscript{2}/V s, which is the highest reported mobility for high In content InGaAs quantum wells to the best of the authors' knowledge. Our results demonstrate that the digital alloy approach can be used to increase the mobility of quantum wells in random alloy ternary materials.

\end{abstract}


\maketitle 

\section{Introduction}

InGaAs quantum wells with high In content (In $\geq$ 75\%) have a smaller electron effective mass, smaller band gap energy, larger \textit{g}-factor \cite{Kosa}, and larger spin-orbit coupling \cite{Jancu2005,Holmes2008} compared to GaAs quantum wells. These properties of InGaAs quantum wells make high electron mobility quantum wells promising materials for topological quantum computing \cite{Oreg2010,Lutchyn2010} and spintronics \cite{Datta1990}. While alloying the GaAs with InAs both enables the tunability of the properties and enhances certain properties of the quantum wells, it introduces random alloy disorder in the material. This intrinsic random alloy disorder leads to alloy disorder scattering of charge carriers and is a limiting factor in high electron mobility InGaAs quantum wells. The current state of the art mobilities in high In content InGaAs quantum wells is $\sim$ 430,000 cm\textsuperscript{2}/V s \cite{Chen2015} and alloy disorder scattering is a major scattering mechanism near the peak mobility \cite{Chen2015,Capotondi2005}. Reducing this intrinsic alloy disorder scattering poses a significant challenge towards improving the mobilities of InGaAs quantum wells.

Digital alloys (DA) are an alternative approach to alloying by growing the material as a short period superlattice  \cite{Kawabe1983} instead of a random alloy. While the properties of digital alloys often closely resemble that of the random alloys, properties such as the band gap \cite{Kawabe1983,Jiang1988,Song2004} and effective mass \cite{Ahmed2022} can vary significantly. Digital alloys have been predominately implemented in optoelectronic devices such as low excess noise avalanche photodetectors \cite{Rockwell2018} and low series resistance distributed Bragg reflectors \cite{Peters1993}. There have been some studies on the transport of heterostructures grown with digital alloys, such as magnetically doped semiconductor quantum wells \cite{Jaroszynski2002}, parabolic quantum wells \cite{Sundaram1991}, and coupled multiple quantum wells \cite{Stormer1986,Baskey1992}. Digital alloys are potentially promising in improving the transport mobility of semiconductor alloys. By growing a digital alloy as an ordered structure comprised of binary semiconductors, which intrinsically do not posses any alloy disorder, it is expected that there is no alloy disorder scattering within the material and the transport mobility will increase \cite{Hoshino2019,Pant2022}. However, a systematic investigation in the effects of digital alloying on the alloy disorder scattering of charge carrier transport has yet to be performed.

In this study, we compare random and digital alloy InGaAs quantum wells grown metamorphically on InP substrates. InAs/GaAs and InAs/InGaAs superlattices are implemented as digital alloy quantum wells. Single subband transport with vanishing magnetoresistance is observed in all of the samples, which is indicative of high quality samples. We find that the InAs/InGaAs DA quantum wells have an enhanced mobility of 545,000 cm\textsuperscript{2}/V s, which is the highest reported mobility for high In content InGaAs quantum well to the best of the authors' knowledge. The mobility of the InAs/GaAs DA quantum well is found to be reduced in comparison to the random alloy sample. By analyzing the scattering mechanisms of the different quantum wells, we attribute the enhancement of the InAs/InGaAs DA mobility due to a reduction in the alloy disorder scattering and the reduction of the InAs/GaAs DA mobility due to increased interfacial roughness and background impurity incorporation. Our results demonstrate that digital alloying is a promising way to improve the mobility of III-V semiconductor alloys, however care must be made with the growth and design of the digital alloys in order to enhance the mobility.

\section{MBE growth, device fabrication, and measurements}

Samples were grown on epi-ready semi-insulating Fe-doped InP (001) wafers (AXT Inc.) in a VG V80H MBE system. The oxide was desorbed by heating the InP wafers under an As\textsubscript{2} flux until the metal rich transition to a $(4 \times 2)$ surface reconstruction was observed by reflection high energy electron diffraction. This transition is used to calibrate the pyrometer temperature to 515 \textsuperscript{o}C. After oxide desoprtion, 100 nm of a nearly lattice matched superlattice of 4 nm In\textsubscript{0.52}Al\textsubscript{0.48}As/ 1 nm In\textsubscript{0.52}Ga\textsubscript{0.48}As was grown at 480 \textsuperscript{o}C. The substrate was then cooled down to 340 \textsuperscript{o}C, and a step graded buffer layer was grown from In\textsubscript{0.575}Al\textsubscript{0.425}As to In\textsubscript{0.825}Al\textsubscript{0.175}As in 50 nm thick steps, where the composition of In would change 2.5\% each step. The buffer layer is then reverse stepped to In\textsubscript{0.75}Al\textsubscript{0.25}As in 50 nm thick steps of 2.5\% In composition changes. 

\begin{figure}[t]
\includegraphics[width=3.375in]{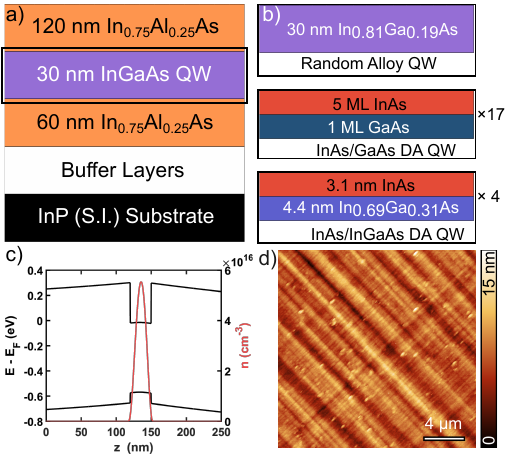}
\caption{\label{fig:1}a) Sample structure schematic, consisting of a 30 nm thick InGaAs quantum well buried 120 nm from the surface. The quantum wells were grown metamorphically on InP subsrates. b) Different samples in the study: a random alloy quantum well, a 17 period InAs/GaAs digital alloy quantum well, and a 4 period InAs/InGaAs digital alloy quantum well. c) Self consistent Schr\"{o}dinger-Poisson simulation of the sample structure. d) Representative AFM micrograph of InAlAs top barrier, cross hatching typical of metamorphic growth is observed.}
\end{figure}

After the completion of the graded buffer layer, the substrate is heated to a pyrometer temperature of 470 \textsuperscript{o}C where a 60 nm In\textsubscript{0.75}Al\textsubscript{0.25}As bottom barrier, a 30 nm In\textsubscript{0.81}Ga\textsubscript{0.19}As quantum well, and a 120 nm In\textsubscript{0.75}Al\textsubscript{0.25}As top barrier are grown. The layer schematic is shown in Fig. \ref{fig:1}(a). The InGaAs quantum well is grown as either a random alloy or a digital alloy, shown schematically in Fig. \ref{fig:1}(b). Two different digital alloy quantum wells with a total thickness of 30 nm are grown: an InAs/GaAs digital alloy quantum well consisting of 17 periods of 5 monolayers of InAs and 1 monolayer of GaAs and an InAs/InGaAs digital alloy quantum well consisting of 4 periods of 3.1 nm of InAs and 4.4 nm of In\textsubscript{0.69}Ga\textsubscript{0.31}As. For the digital alloy samples, a 1 second growth interrupt between the layers was included to improve the ordering at the interfaces. The GaAs layer thickness in the InAs/GaAs digital alloy could not exceed 1 monolayer, growth thicker than 1 monolayer resulted in the relaxation of the GaAs layer. 

The quantum wells are in electron accumulation despite the entire heterostructure being unintentionally doped. This result is consistent with 1D Schr\"{o}dinger-Poisson calculations of the band structure, shown in Fig. \ref{fig:1}(c), performed using the methods described in \cite{Tan1990} and prior reports on the growth of undoped high In content InGaAs quantum wells \cite{Capotondi2004,Simmonds2007}. 

The surface morphology of the InAlAs top barrier is shown in the atomic force micrograph shown in Fig. \ref{fig:1}(d). A cross-hatched morphology typical of metamorphic growth is observed, with the rms roughness of the InAlAs top barrier over a 400 $\mu$m\textsuperscript{2} area typically 1-2 nm.

Mesas for Hall bars were fabricated using wet chemical etching with a mixture of H\textsubscript{2}SO\textsubscript{4}:H\textsubscript{2}O\textsubscript{2}:H\textsubscript{2}O (1:8:120). Following mesa definition, NiGeAu contacts were deposited and subsequently annealled at 450 \textsuperscript{o}C for 2 minutes to form ohmic contacts to the quantum well. A 30 nm AlO\textsubscript{x} gate dielectric was deposited with atomic layer deposition. Finally, a Ti/Au top gate was deposted on the Hall bar. The ratio of the Hall bar width to the spacing of the arms was 9.25. Magnetotransport measurements were performed in a Quantum Design Physical Property Measurement System at 2 K using standard low frequency AC lock-in measurements, with an AC current of 0.5 $\mu$A. The two-dimensional electron gas (2DEG) carrier density was modulated by applying a voltage to the top gate electrode.

\begin{figure*}[t]
\includegraphics[width=6.750in]{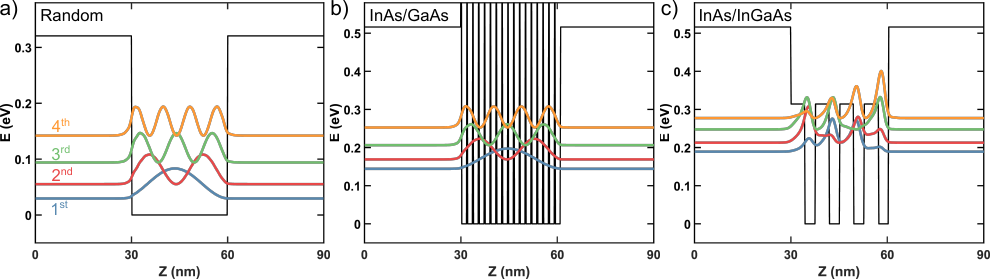}
\caption{\label{fig:2}Plots of the conduction band edges and the $|\psi|^{2}$ of the first four conduction band subbands of the: a) random alloy, b) InAs/GaAs DA, and c) InAs/InGaAs DA quantum wells calculated with the $k\cdot p$ method.}
\end{figure*}

\section{Results and Discussion}
\subsection{Quantum well band structure}
The band structure of the random and digital alloy quantum wells was computed using the $k\cdot p$ method for heterostructures with the envelope function approximation, using the methods described in \cite{Vurgaftman2021}. The first four conduction band subband energies and their associated probability densities ($|\psi|^{2}$) for the three different samples are plotted in Fig. \ref{fig:2}. The $|\psi|^{2}$ of the random alloy and the InAs/GaAs DA quantum wells are sinusoidal-like envelop functions over the channel of the quantum well. However, for the $|\psi|^{2}$ in the InAs/InGaAs digital alloy, the $|\psi|^{2}$ are the sinusoidal-like envelope functions of the quantum well with peaks in the InAs layers and troughs in the InGaAs regions. Although the InAs/InGaAs digital alloy still has alloy disorder present through the use of the InGaAs layers, the overall alloy disorder scattering may be reduced by having the electron preferring to reside in the InAs regions of the digital alloy. The band parameters of the quantum wells are summarized in Table \ref{tab:1}, which contains the predicted energy gap of the quantum wells, the subband spacing between the 1\textsuperscript{st} and 2\textsuperscript{nd} electron subbands, the 1\textsuperscript{st} subband effective mass ($m^{*}$), and the nonparabilicity factor ($\alpha$). $\alpha$ is defined with: $E(1+\alpha E) = \frac{\hbar^{2}k^{2}}{2m^{*}} $, where $E$ is the Fermi energy and $k$ is the Fermi wavevector.

From the $k\cdot p$ calculations, it is expected that subbands form in the digital alloy and random alloy quantum wells, with the digital alloy quantum wells having a slightly smaller subband spacing than the random alloy quantum wells. The predicted effective masses and nonparabilicity of the quantum wells is not expected to change significantly. However, it is predicted that the InAs/GaAs DA quantum well to have a slightly smaller effective mass compared to the InAs/InGaAs DA and the random alloy quantum wells. The predicted results for the InAs/GaAs DA may be incorrect, as the $k\cdot p$ method can be less accurate in simulation of monolayer thick heterostructures, which is discussed in depth in \cite{Wood1996}.

\begin{table}[b]
\caption{\label{tab:1}
$k \cdot p$ results including band gap energy ($E_g$), subband spacing ($E_2 - E_1$), 1\textsubscript{st} subband effective mass ($m^*$), and effective mass nonparabilicity ($\alpha$).}
\begin{ruledtabular}
\begin{tabular}{ l c c c c}
  & $E_{g}$ &  $E_{2} - E_{1}$ & $m^{*}$ & $\alpha$ \\ 
 Sample & ($meV$) & ($meV$) & ($m_{e}$) & \\ 
\hline
 Random Alloy & 573 & 25.7 & $0.0347 $ & 1.955 \\  
 InAs/GaAs DA & 566 & 24.7 & $0.0334 $ & 2.134 \\
 InAs/InGaAs DA & 551 & 23.6 & $0.0345 $ & 2.324 \\
\end{tabular}
\end{ruledtabular}
\end{table}


\subsection{Magnetotransport}
\begin{figure}[b]
\includegraphics[width=3.375in]{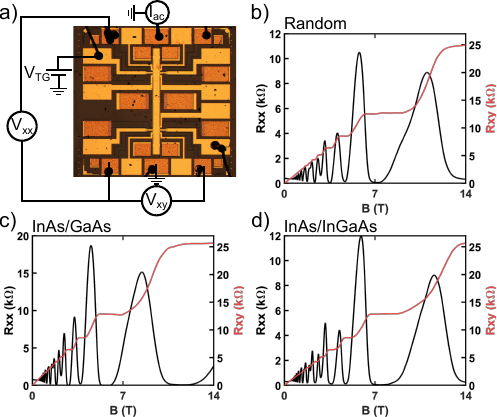}
\caption{\label{fig:3}a) Optical image of the Hall bar and schematic sample wiring. 2 K magnetotransport of the: b) random alloy sample at $n_{2DEG}$ $=$ $3.52 \times 10^{11}$ cm\textsuperscript{-2}. c) InAs/GaAs DA sample at $n_{2DEG}$ $=$ $2.78 \times 10^{11}$ cm\textsuperscript{-2}. d) InAs/InGaAs DA sample at $n_{2DEG}$ $=$ $3.60 \times 10^{11}$ cm\textsuperscript{-2}.}
\end{figure}

\begin{figure*}[t]
\includegraphics[width=6.750in]{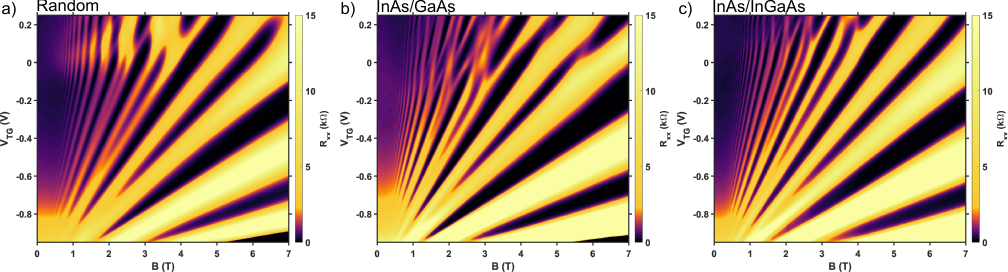}
\caption{\label{fig:4}R\textsubscript{xx} at different top-gate voltages and magnetic fields for: a) random alloy, b) InAs/GaAs DA, and c) InAs/InGaAs DA.}
\end{figure*}

An optical image of a Hall bar used in this study with the schematic wiring diagram is shown in Fig. \ref{fig:3}(a). The R\textsubscript{xx} and R\textsubscript{xy} from B = 0 T to B = 14 T for the random and digital alloy quantum wells is shown in Fig. \ref{fig:3}(b-d), with $n_{2DEG}$ $=$ $3.52 \times 10^{11}$, $2.78 \times 10^{11}$, and $3.60 \times 10^{11}$ cm\textsuperscript{-2} for the random alloy, InAs/GaAs DA, and InAs/InGaAs DA, respectively. $n_{2DEG}$ is the sheet electron density extracted from the Hall measurements. The integer quantum Hall effect is observed along with vanishing magnetoresistance for all samples. In both of the digital alloy quantum wells, single band transport is observed and the transport does not resemble the coupled multiple quantum well transport observed in GaAs/AlAs digital alloy quantum wells \cite{Stormer1986,Baskey1992}. We propose this difference is due to stronger coupling between the sublattices of the digital alloy compared to the GaAs/AlAs. Plots of the magnetoresistance at various gate voltages for the different samples are shown in Fig. \ref{fig:4}. At lower (more negative) gate voltages, a single set of Shubnikov–de Haas oscillations is observed in all of the quantum wells. At higher gate voltages, the second subband becomes occupied and an additional set of oscillations is observed for all of the samples. These results demonstrate the formation of a high quality two-dimensional electron gas devoid of parallel conduction in the random alloy quantum well and the digital alloy quantum wells.

\begin{table*}
\caption{\label{tab:2} Sample parameters including the peak mobility, carrier density at the peak mobility, the effective mass from thermal damping, the estimated Fermi energy at the peak mobility, and the carrier density a V\textsubscript{TG} = 0 V.}
\begin{ruledtabular}
\begin{tabular}{l c c c c c}
& $\mu_{max}$ & $n_{2DEG}$ at $\mu_{max}$ & $m^{*}$ & $\varepsilon_{f}$ at $\mu_{max}$ & $n_{2DEG}$ at V\textsubscript{TG} = 0 V\\
 Sample & ($10^{3}$ cm\textsuperscript{2}/Vs) & ($10^{11}$ cm\textsuperscript{-2}) & ($m_{e}$) & (meV) &  ($10^{11}$ cm\textsuperscript{-2})\\ 
\colrule
 Random Alloy & 460 & 3.13 & $0.030 \pm 0.001$ & 25.0 &  3.81\\  
 InAs/GaAs DA & 281 & 2.59 & $0.030\pm 0.001$ & 20.7 &  3.73\\
 InAs/InGaAs DA & 545 & 3.37 & $0.033\pm 0.002$ & 24.5 & 3.89\\
\end{tabular}
\end{ruledtabular}
\end{table*}

\begin{figure}[b]
\includegraphics[width=3.375in]{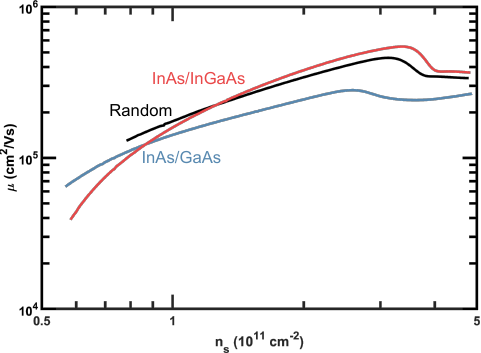}
\caption{\label{fig:5} The dependence of mobility on sheet carrier density for the three different samples. Carrier density was varied by modifying the top gate voltage.}
\end{figure}

From the magnetotransport at various gate voltages, the mobility at different carrier densities can be determined and are shown in Fig. \ref{fig:5} for the different quantum wells. The random alloy digital alloy quantum well has a peak mobility of 460,000 cm\textsuperscript{2}/Vs at a $n_{2DEG}$ of $3.13 \times 10^{11}$ cm\textsuperscript{-2}, the InAs/GaAs DA quantum well has a peak mobility of 281,000 cm\textsuperscript{2}/Vs at a $n_{2DEG}$ of $2.59 \times 10^{11}$ cm\textsuperscript{-2}, and the InAs/InGaAs DA quantum well has a peak mobility of 545,000 cm\textsuperscript{2}/Vs at a $n_{2DEG}$ of $3.37\times 10^{11}$ cm\textsuperscript{-2}. The mobility of the random alloy quantum well is slightly higher than the mobility of 430,000 cm\textsuperscript{2}/Vs reported in the current state of the art high In content InGaAs quantum wells \cite{Chen2015}. The InAs/InGaAs DA quantum well has a higher peak mobility at a higher carrier density than the random alloy quantum well, while the InAs/GaAs DA samples has a lower peak mobility at a lower carrier density than the random alloy quantum well. From the results shown in Fig. \ref{fig:4}, the decrease in mobility after the peak mobility in all of the samples is likely due to intersubband scattering. The electron effective mass is extracted from the thermal damping on the Shubnikov–de Haas oscillations, and these effective masses were used to estimate the Fermi energy at the peak mobility and to estimate the subband splitting. These results are summarized in Table \ref{tab:2}. All of the samples have smaller effective masses than the predicted effective masses from the $k\cdot p$ calculations, and in the case of the random alloy and the InAs/GaAs DA quantum wells, they have smaller effective masses than 0.0314 $m_e$, which is the expected bulk effective mass of InGaAs accounting for strain effects calculated using $k\cdot p$ \cite{Vurgaftman2021}. This reduction in effective mass is likely due to electron-electron correlations, and has been observed before in GaAs quantum wells \cite{Hatke2013}. The subband energies estimated from the effective mass and the peak carrier density for the random and the InAs/InGaAs DA are within $\sim 1$ meV of the predicted subband energies, and the higher carrier density at the peak mobility for the InAs/InGaAs quantum well is due to the larger effective mass of the InAs/InGaAs sample. However, the InAs/GaAs DA quantum well subband energy is significantly different than the predicted, and may be due to the aforementioned potential inaccuracies of the $k\cdot p$ method for monolayer heterostructures.

\subsection{Scattering mechanisms}

\begin{table}[b]
\caption{\label{tab:3} Mobility modeling parameters including alloy disorder potential ($dV$), background impurity density ($n_{BI}$), interface roughness ($\Delta$), and correlation length ($\Lambda$).}
\begin{ruledtabular}
\begin{tabular}{ l c c c c}
  & $dV$ &  $n_{BI}$ & $\Delta$ & $\Lambda$ \\ 
 Sample & (eV)  & (cm\textsuperscript{-3}) & (\r{A}) & (nm)  \\ 
\hline
 Random Alloy & 0.44 & $1.5\times10^{15}$  & 6 & 13 \\  
 InAs/GaAs DA & 0.44 & $2.4\times10^{15}$  & 21 & 3\\
 InAs/InGaAs DA & 0.38 & $1.6\times10^{15}$  & 6 & 13\\
\end{tabular}
\end{ruledtabular}
\end{table}

\begin{figure*}[t]
\includegraphics[width=6.750in]{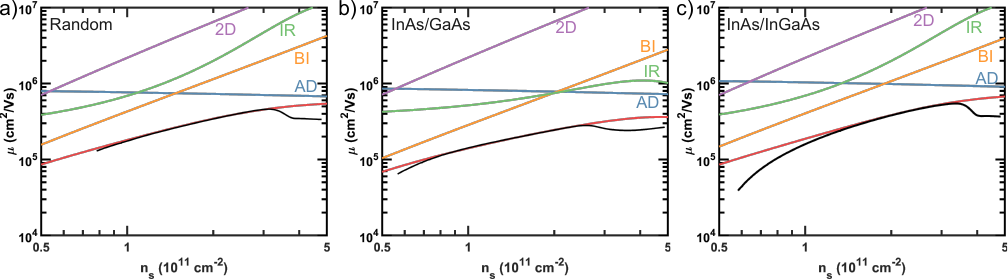}
\caption{\label{fig:6} Experimentally measured mobility and calculated mobility at various sheet carrier densities for the: a) random alloy, b) InAs/GaAs DA, and c) InAs/InGaAs quantum wells. The total mobility and the individual mobilities for the different scattering mechanisms including: interfacial roughness scattering (IR), two-dimensional remote impurity scattering (2D), background impurity scattering (BI), and alloy disorder scattering scattering (AD) are plotted.}
\end{figure*}

To understand the differences in the mobility and the scattering mechanisms for the different samples, the mobilities were modeled using a calculation for the elastic scattering time of electrons in the two-dimensional electron gas \cite{Stern1967}. The calculations accounted for background impurity scattering \cite{Gold1987}, interfacial roughness scattering \cite{Sakaki1987,Jana2011}, remote impurity scattering \cite{Gold1987}, and alloy disorder scattering \cite{Bastard1983}. Zero temperature was assumed and intersubband scattering was not considered. The exact details of the calculations are given in the appendix. The remote impurity scattering was treated as a fixed sheet charge of $2\times 10^{12}$ cm\textsuperscript{-2} at the top surface of the quantum well, which is an overestimate of the surface state density of InAlAs \cite{Hwang1994, Chou1998, Chang1999, Hwang2001}. However, the exact surface state density of the remote impurities at the surface is found to be not important, as the contribution of remote impurity scattering from the top surface is found to have negligible effect on the mobility of the quantum wells due to the depth of the quantum well from the surface. The parameters for the other scattering mechanism were varied to obtain good agreement with the experimental data, and are given in Table \ref{tab:3}. The transport lifetimes due to the different scattering were calculated independently, and the total mobility was computed using Matthiessen's rule.

The mobilities of the samples at different carrier densities and the calculated mobilities are shown in Fig. \ref{fig:6}. The calculated total mobility and the mobility for the different scattering mechanisms are also shown. There is good agreement between the calculated mobility and the measured mobility for all of the samples at higher carrier densities until the peak mobility is reached, where the measured mobility begins to decrease due to intersubband scattering between the 1\textsuperscript{st} and 2\textsuperscript{nd} subbands, which is not accounted for with the mobility calculations. In all of the samples, two distinct regimes can be observed, alloy disorder scattering dominates at higher carrier densities and background impurity scattering dominates at lower carrier densities. This behavior is similar to what has been observed in prior studies of InGaAs quantum wells \cite{Capotondi2005,Chen2015}. The measured mobility begins to diverge from the calculated and is lower than what the calculations would predict at carrier densities lower than $\sim 1\times 10^{10}$ cm\textsuperscript{-2}. This deviation in mobility is likely a metal-insulator transition due to localization of charge carriers. Similar metal-insulator transitions have been observed before in GaAs \cite{Gold1991}, AlAs \cite{Gold2008}, and Si quantum wells \cite{Gold2010}.

A higher background impurity density is required to accurately model the mobilities of the InAs/GaAs ($2.4 \times 10^{15}$ cm\textsuperscript{-3}) and the InAs/InGaAs ($1.6 \times 10^{15}$ cm\textsuperscript{-3}) digital alloy quantum wells than the random alloy quantum well ($1.5 \times 10^{15}$ cm\textsuperscript{-3}). The background impurity densities extracted from the modeling are similar to what has been previously reported in InGaAs quantum wells \cite{Capotondi2004,Capotondi2005,Chen2015}. The InAs/GaAs DA has more interfaces, with the 17 periods of InAs/GaAs, than the InAs/InGaAs DA, with 4 periods, and has a higher estimated background impurity density than the InAs/InGaAs DA. This trend suggests that the higher background impurity density is due to impurity incorporation at the interfaces of the digital alloy during growth, which has been observed to occur at the interfaces of superlattices grown by MBE \cite{Achtnich1987,Yoon1993}. The origin of these additional impurities could be from the brief growth interrupt during the growth of the digital alloys. From these results, it is desirable to reduce the number of periods in the superlattice to reduce the background impurity density and therefore the background impurity scattering for the growth of high mobility samples.

The interfacial roughness in the random and InAs/InGaAs digital alloy is estimated from the model to be 6 \r{A}. This value is smaller than the roughness measured by atomic force microscopy of the InAlAs top barrier, which we attribute due to the InGaAs layers being smoother than the InAlAs layers at the growth temperature.  The interfacial roughness of the random and InAs/InGaAs DA quantum wells is comparable to prior studies of the interfacial roughness of InGaAs quantum wells \cite{Chen2015}. However, the modeled interfacial roughness of the InAs/GaAs is significantly higher, with a roughness of 21 \r{A}. The increased roughness becomes a significant scattering mechanism at higher carrier densities in the InAs/GaAs digital alloy. The increased roughness is likely due to the growth of rougher GaAs islands as the GaAs layer in the InAs/GaAs digital alloy. GaAs has been reported to grow as islands on InAs layers \cite{Schowengerdt1988}. These results demonstrate that the optimal growth temperature of the constituent layers of the digital alloy is an important consideration, and can result in rougher growth and enhanced interfacial roughness scattering.

In order to accurately model the experimental data, the alloy disorder potential in the random alloy quantum well is estimated to be 0.44 $eV$. This value is larger than the 0.3 $eV$ reported previously in InGaAs quantum wells \cite{Chen2015}, but closer to the expected value of 0.5 $eV$ of InGaAs \cite{Ferry1978}. We find that the same alloy disorder potential of 0.44 $eV$ is required to obtain a good agreement to the experimental InAs/GaAs DA mobility curves, which we propose is due to significant intermixing in the InAs/GaAs layers creating alloy disorder. In the InAs/InGaAs digital alloy, the alloy disorder potential is estimated to be 0.38 $eV$ from the modeling, this alloy disorder potential is smaller than the alloy disorder of the random alloy. The reduction in the alloy disorder potential is likely due to the electrons preferentially residing in the InAs regions of the digital alloy, as seen in the $k\cdot p$ results in Fig. \ref{fig:2}c, and as a consequence the electron do not spend as much time in the regions of the quantum well that posses alloy disorder. The overall increase in mobility observed in the InAs/InGaAs DA quatnum well over the random alloy qauntum well is due to this reduction in alloy disorder scattering. These results as a whole suggest that digital alloying can be used to reduce alloy disorder scattering in quantum wells and the importance of the heterostructure design and the interfaces in the mobility of digital alloy quantum wells.

\section{Conclusions}
In summary, we have demonstrated in high electron mobility InGaAs quantum wells, digital alloys can be used as an alternative to random alloys in semiconductor quantum wells and increase the quantum well mobility by reducing the alloy disorder scattering. Single subband transport, the interger quantum Hall effect, and the absense of parallel conduction are observed in the random and digital alloys. However, increased background impurity density is observed in the digital alloys, likely due to incorporation of impurities at the interfaces. The results indicate the digital alloy should be grown with as few periods as possible to improve the mobility of the digital alloy quantum wells over that of the random alloy. Additionally, the quality of the interfaces and the growth conditions are important, otherwise alloy disorder and enhanced interfacial roughness scattering can be introduced into the digital alloy quantum wells and subsequently reduce the mobility. We believe that our results should be generally applicable to other semiconductor alloys, and can be used to increase the mobility in other material systems.

\begin{acknowledgments}
This growth of the quantum wells was supported by University of California Multiple Campus Award No. 00023195. Modeling efforts of the quantum wells was supported by the Department of Energy under award No. DE-SC0019274. We acknowledge the use of shared facilities of the NSF Materials Research Science and Engineering Center (MRSEC) at the University of California Santa Barbara (Grant No. DMR 2308708) and the Nanotech UCSB Nanofabrication Facility.
\end{acknowledgments}

\appendix
\section{Mobility modeling}

The InGaAs quantum well with InAlAs barriers envelope wave function ($\phi (z)$) over the InGaAs thickness $L$ is approximated by: $$\phi (z) = (\frac{2}{L})^{1/2} sin(\frac{\pi z}{L})$$ for $0 \leq z \leq L$ and 0 for all other $z$. Only the lowest occupied subband is treated. The mobility is defined as: $\mu = e\tau/m^{*}$, where $e$ is the elementary charge, $\tau$ is the scattering time, and $m^{*}$ is the effective mass.

From the treatment by Stern and Howard \cite{Stern1967} using the Born approximation, the scattering time for a scattering mechanism ($i$) at a given Fermi wavevector ($k_{f}$) is defined as: $$\frac{\hbar}{\tau_{i}} = \frac{1}{2 \pi \varepsilon_{f}}\int_{0}^{2k_{f}} dq \frac{q^2}{\sqrt{4k_{f}^2 - q^2}} \frac{\langle|U_{i}(q)|^2 \rangle}{\epsilon (q)^2}$$ where $q$ is the scattering vector, $\varepsilon_{f}$ is the Fermi energy computed using a nonparabolic effective mass, $\hbar$ is the reduced Planck constant, $\langle|U_{i}(q)|^2 \rangle$ is the averaged scattering potential associated with the particular scattering mechanism, and $\epsilon(q)$ is the dielectric function from the random phase approximation. The dielectric function is defined as: $$\epsilon(q) = 1 + \frac{q_{TF}}{q}[1-G(q)]F_{C}(q)$$ where $q_{TF}$ is the Thomas-Fermi screening wavenumber $q_{TF} = m^{*} e^{2} / 4 \pi \epsilon_{L} \hbar^{2}$ with the dielectric constant $\epsilon_L$, $G(q)$ is the generalized form of the local field correction \cite{AmaliaYuniaRahmawati2020}, and $F_C(q)$ is the form factor of the Coulomb interaction of the finite width of the quantum well, defined as: $$F_C(q) = \int_{- \infty}^{+ \infty}dz |\phi(z)|^{2}\int_{- \infty}^{+ \infty}dz'|\phi(z')|^{2}exp(-q|z-z'|)$$

The scattering potential for interface roughness is \cite{Sakaki1987} of the form: $$\langle|U_{IR}(q)|^2 \rangle = \pi F_{IR}^2 \Delta^2 \Lambda^2 exp(-(q\Lambda)^2/4)$$ where $\Delta$ is the interfacial roughness, $\Lambda$ is the correlation length, and $F_{IR}$ is a function that is modified following the treatment by \cite{Jana2011} to include the shift of the ground state by a electric field $E$: $$F_{IR} = -(\frac{\hbar^2 \pi^2}{m^* L^3} + 96 (\frac{2}{3\pi})^6 \frac{e^2 m^* L^3 E^2}{\hbar^2})$$

Alloy disorder within the In\textsubscript{$1-x$}Ga\textsubscript{$x$}As quantum well with an alloy disorder potential of $dV$ leads to a scattering potential \cite{Bastard1983} of:
$$\langle|U_{AD}(q)|^2 \rangle = x(1-x) dV^2 \frac{\sqrt{3} \pi a^3}{16} \frac{3L}{2}$$

Two-dimensional remote ionized impurities with an charged impurity density of $n_{j}$ lead to a scattering potential of the form \cite{Gold1987}: $$\langle|U_{2D}(q)|^2 \rangle = (\frac{2 \pi e^2}{\epsilon_L}\frac{1}{q})^2 n_j F_{2D}(q,z_j)^2$$ $F_{2D}(q,z_j)$ is the form factor accounting for the distance between the charged impurity layer and the quantum well: $$F_{2D}(q,z_j) = \int_{-\infty}^{+\infty} dz |\phi(z)|^{2} exp(-q|z-z_{j}|)$$

The scattering potential of homogeneously distributed charged background impurities \cite{Gold1987} is: $$\langle|U_{BI}(q)|^2 \rangle = (\frac{2 \pi e^2}{\epsilon_L}\frac{1}{q})^2 N_B L F_{BI}(q)$$ where $N_B$ is the background impurity density and $F_BI(q)$ is the form factor for background impurities, defined as: $$F_{BI}(q) = \frac{1}{L}\int_{-\infty}^{+\infty} dz_{j} F_{2D}(q,z_j)^2$$ 

The total scattering time was determined from the individual scattering times and applying Matthiessen's rule: $$1/\tau = 1/\tau_{IR} + 1/\tau_{AD} + 1/\tau_{2D} + 1/\tau_{BI}$$

\bibliography{References}

\end{document}